\documentclass[12pt,centertags,reqno]{amsart}

\usepackage[foot]{amsaddr}
\usepackage{latexsym}
\usepackage[english]{babel}
\usepackage[T1]{fontenc}

\usepackage[utf8]{inputenc}
\usepackage[numbers]{natbib}
\usepackage{amssymb,amsmath}
\usepackage{fancyhdr}
\usepackage{url}
\usepackage{hyperref}
\usepackage{verbatim}
\usepackage{leftidx}
\usepackage{bm}
\usepackage{color,graphicx}
\usepackage{tikz}
\usepackage{bbm}

\usepackage{mathrsfs, mathtools}
\usepackage{stmaryrd}

\usepackage{marvosym}
\mathtoolsset{showonlyrefs}

\usepackage{appendix}

\usepackage{soul}

\usepackage{enumerate} 
\usepackage{enumitem} 

\numberwithin{equation}{section} \makeatletter
\renewcommand{\subsection}{\@startsection
{subsection}{2}{0mm}{\baselineskip}{-0.25cm}
{\normalfont\normalsize\bf}} \makeatother

\newtheorem{theorem}{Theorem}[section]
\newtheorem{lemma}[theorem]{Lemma}

\newtheorem{definition}[theorem]{Definition}
\newtheorem{remark}[theorem]{Remark}
\newtheorem{proposition}[theorem]{Proposition}

\def \A {\mathcal A}

\def \F {\mathcal F}
\def \G {\mathcal G}

\def \L {\mathcal L}

\def \P {\mathbf P}
\def \Q {\mathbf Q}
\def \I {{\mathbf 1}}

\def \R {\mathbb R}
\def \bF {\mathbb F}
\def \bG {\mathbb G}

\def \bN {\mathbb N}

\newcommand{\ud}{\mathrm d}

\newcommand{\esp}[2][\mathbb E] {#1\left[#2\right]}

\begin{document}

\author[C.~Ceci]{Claudia  Ceci}
\address{Claudia  Ceci, Department MEMOTEF,
University of Rome Sapienza, Via del Castro Laurenziano, 9,
Rome, Italy.}\email{claudia.ceci@uniroma1.it}
\author[K.~Colaneri]{Katia Colaneri}
\address{Katia Colaneri, Department of Economics and Finance, University of Rome Tor Vergata, Via Columbia 2, 00133 Rome, Italy.}\email{katia.colaneri@uniroma2.it}

\title[Portfolio and reinsurance optimization under unknown market price of risk]{Portfolio and reinsurance optimization under unknown market price of risk}

\date{\today}

\begin{abstract}
We investigate the optimal investment-reinsurance problem for insurance company with partial information on the market price of the risk. Through the use of filtering techniques we convert the original optimization problem involving different filtrations, into an equivalent stochastic control problem  under the observation filtration only, the so-called {\it separated problem}. The Markovian structure of the separated problem allows us to apply a classical approach to stochastic optimization based on the Hamilton-Jacobi-Bellman equation, and to provide explicit formulas for the value function and the optimal investment-reinsurance strategy. We finally discuss some  comparisons between the optimal strategies pursued by a partially informed insurer and that followed by a fully informed insurer, and we evaluate the value of information using the idea of indifference pricing. These results are also supported by numerical experiments.
\end{abstract}

\maketitle

{\bf Keywords}: Optimal investment and reinsurance; stochastic control; Kalman filter.

{\bf JEL Classification}: G11, G22, C61

{\bf AMS Classification}: 93E20, 93E11, 60G55, 60J60.

\section{Introduction}

Optimal reinsurance-investment is a key problem within the actuarial literature. Reinsurance agreements serve as key instruments in risk management, enabling primary insurers to transfer a portion of their risk exposure to some third party. Compliance with these contracts is mandated by certain financial regulations, as, for instance, the Solvency II regulatory framework within the European Union. Engaging with a  reinsurance contract permits insurers to mitigate unfavorable claim scenarios and broaden risk diversification, bringing potentially to higher returns on the cash flows of insurance portfolios. 
At the same time, insurance firms allocate their resources across different assets like stocks, bonds, real estate, and alternative investments. The extra yield from these investments, provides additional sources of revenues that complement the earnings from insurance premiums, and simultaneously strengthens the financial stability of the company. Furthermore, sound investment strategies are crucial for maintaining sufficient capital levels, which is fundamental for  insurers to meet the solvency regulations. 

The academic literature has investigated optimal reinsurance and investment problems under different modeling frameworks and using a broad variety of criteria, as, for example, the maximization of  expected utility from terminal wealth, mean-variance objectives or the minimization of ruin probability (see, for example \citet{Schmidli2002}). In this paper we consider the approach of the utility maximization and solve the optimal reinsurance and investment problem via the dynamic programming approach. We consider a framework where the market is affected by a hidden factor process that cannot be directly measured and formulate a stochastic control problem under partial information. We adopt a widely accepted setup where the market price of risk is unobservable, and we estimate it using techniques from stochastic filtering. Our ultimate goal is to characterise explicitly the investment strategy, the reinsurance agreement, and the value process. Therefore, we consider a simplified setting where the market dynamics are driven by geometric Brownian motion. Although simplistic,  this choice has the advantage to provide  solutions in closed form. Moreover, it makes it possible to discuss the comparison with a setup of full information on all state variables.

\subsection{Literature review}

As mentioned at the beginning of the introduction, the problem of investment and reinsurance is standard in the actuarial setup. Classical textbooks and pioneering articles on the topic are \citet{Schmidli2002}, \cite{Schmidli2008}, \cite{Schmidli2018}, \citet{Irgens2004}, \citet{Liu2009}. 
The classical setting is characterised by the assumptions of  full information on all processes that govern the dynamics of the financial and the insurance markets, which, in addition are assumed to be independent. The more recent literature has addressed the problem under more involved frameworks where the assumption of independence is dropped by taking into account, for instance common shock between risky asset prices and the loss process (see, e.g. \citet{Liang2016} and \citet{Liang2018}, or \citet{ceci2022IME}).  
There are also results on optimal investment and reinsurance under partial information. Cases where unobservable factors only act on the financial market are discussed, for instance, in \citet{Peng2013}, \citet{LiQiao2015}, \citet{Liang-Song2015}, \citet{BiCaoZeng2021}. Few other studies consider partial information on the actuarial market. For example, \citet{LiangBay} introduce a Markov modulated risk model with claim arrival intensity and claim size distribution driven by an unobservable finite-state continuous time Markov chain. They solve the optimization by means of the Hamilton Jacobi Bellman equation (in short HJB). A partial information setup with unobservable claim arrival intensity is also considered in \citet{bauerle2021robust}. They use a Bayesian approach allowing for learning unknown model parameters, and study the optimization problem under exponential
utility preferences. 
In \citet{brachetta2020IME} Markovian unobservable stochastic factor with a more general dynamics is  considered. There, the main issue is related to infinite dimensionality of the filter, hence the problem is addressed via Backward Stochastic Differential Equations (BSDEs). With a similar methodology, \citet{BCCS2023} study a partially observable risk model which combines Hawkes and Cox with shot noise intensity processes and exhibits cluster features.
The main drawback of these papers stand in the fact that no explicit solution is derived, which makes it complicated to discuss comparison results.

Therefore, in this paper, we deliberately consider a simplified setting, with the aim to provide explicit expressions for the optimal reinsurance and investment strategy under partial information. We also seek to characterise the value function in closed form and discuss comparisons between optimal strategies  under complete and partial information, both analytically and numerically.

One key issue when dealing with optimization problems under partial information is the existence of optimal controls. \citet{fleming1982optimal}, \citet{el1988existence} and \citet{elliott1992partially}, among the first ones, suggested to address the problem by introducing a new control problem called the \emph{separated problem}, which is proven to be equivalent to the original control problem (see, e.g. \cite{fleming1982optimal} for more details).  In short, this auxiliary problem corresponds to the control problem where  unobservable factors are replaced by their conditional distribution and all state processes become observable.  To solve the separated problem, different techniques can be applied, mainly depending on the properties of the filter. 
In our framework the dynamic programming approach via the (finite dimensional) HJB equation can be used, and allows to characterise both the optimal control process and the value function in terms of its stochastic representation or in a quasi-explicit way.

The paper is structured as follows. In Section \ref{sec:problem}  we present the market model under partial information and formulate the optimal investment and reinsurance problem for an insurer with exponential utility preferences. 
In Section \ref{sec:reduction}, we focus on the application of the Kalman-Bucy filter to convert the optimization problem under partial information to an equivalent optimization problem under the observation filtration. In Section \ref{sec:HJB} we solve the optimization problem using the Hamilton Jacobi Bellman approach and provide a Verification theorem. In Section \ref{sec:verification} we discuss conditions to ensure that the optimal strategy falls in the class of admissible strategies. 
Finally, in Section \ref{sec:comparison} we present some comparison results where we employ  analytical techniques for special values of the parameters and numerical experiments in more general cases. In particular, we borrow the definition of indifference price to evaluate the value of information. Appendix \ref{app:technical} collects the proofs of some of the technical results stated in the body of the paper.

\section{Problem formulation}\label{sec:problem} 

We consider a combined financial-insurance market model described as follows. Let $T<\infty$ be a fixed time horizon and let $(\Omega, \G, \P)$ be a probability space with a complete and right continuous filtration  $\bG = \{ \G_t\} _{t \in [0,T]}$, such that $\mathcal G_T=\mathcal G$. In the sequel we refer to $\bG$ as the global filtration and assume that all processes defined below are $\bG$-adapted, if not otherwise specified. 

\subsection{The insurance setup}
The insurance company faces claims that are mathematically described by the classical Cram\'er-Lundberg model (see, e.g. standard textbooks as  \citet{Grandell, Schmidli2018} for more details).
We let $\{T_n\}_{n \ge 0}$ be sequence of random times that represent the times at which claim events occur. We assume that inter-arrival times, $T_{i+1}-T_i$, are exponentially distributed, with parameter $\lambda >0$ and denote by $N = \{N_t\}_{t \in [0,T]}$ the process that counts the number of claim events. $N$ is a Poisson process with constant intensity $\lambda$. We let $\{Z_n\}_{n \geq 1}$ be a sequence of independent, identically distributed,  positive random variables that indicate the claim amounts. We denote by $F_Z : (0, + \infty) \to \R$ the cumulative distribution function of $Z_i$ and assume that $\mathbb{E}[e^{aZ_i}] = \int_{\R^+}e^{az} F_Z(\ud z) < \infty$, for any $a >0$ \footnote{This is a tecnical assumption that is required later to ensure the well-posedness of the optimization problem.}. We also assume that $\{N_t\}_{t \in [0,T]}$ and $\{Z_n\}_{n \geq 1}$ are independent of each other.  Then we define the cumulative claim process as
\begin{equation}
C_t = \sum_{n =1}^{N_t} Z_n, \quad t \in [0,T].
\end{equation}
This process keeps track of the total loss of the insurance company due to claims. It may be convenient later on, to consider the representation of the cumulative claim process in terms of the the natural integer-valued measure
$$m(\ud t, \ud z) = \sum_{n \geq 1}\delta_{\{T_n, Z_n\}}(\ud t, \ud z)\I_{T_n\leq T}, 
$$
so that
\begin{equation}
    C_t=\int_0^t \int_{\R^+} z m(\ud s, \ud z),  \quad t \in [0,T].
\end{equation}
The measure $m(\ud t, \ud z)$ is a Poisson random measure with the compensator given by $\lambda F_Z(\ud z) \ud t$. \footnote{In particular, it holds that for any $\bG$-predictable random field $H(t,z)$ such that $\mathbb{E}\left[\int_0^T\int_{\mathbb{R}^+} H(t,z) F_{Z}(\ud z) \ud t \right]<\infty$, the process $$\int_0^t\int_{\mathbb{R^+}}H(s, z) \left(m (\ud s, \ud z)-\lambda F_{Z}(\ud z)\ud s\right)$$ is a martingale with respect to the probability $\P$ and the filtration $\bG$.} The insurance premiums are modeled via an insurance premium rate $c>0$.   

The insurance company is allowed to buy proportional reinsurance with a retention level $u_t\in [0,1]$, that may potentially be dynamic. At any time $t$, the value $u_t \cdot 100 \%$ denotes the percentage of total claims that are retained by the insurance company, and consequently $(1-u_t)\cdot 100 \%$ is the percentage of claims that are covered by the reinsurer. Therefore, the aggregate loss which is covered by the insurer is given by
$$
C^u_t = \sum_{n=1}^{N_t} u_{T_n} Z_n = \int_0^t \int_{\R^+} u_s  z m(\ud s, \ud z),  \quad t \in [0,T].
$$
The insurer pays a reinsurance premium, in the form of a rate $q(u_t)>0$.

Hence, the surplus of the insurer is described by the process $R^u$, which follows the SDE 
\[
dR^u_t = \bigl(c-q(u_t)\bigl)\,dt - dC^u_t, \quad R_0^u = R_0 \in \mathbb R^+,
\]
for some initial capital $R_0$. 

\begin{remark}[Classical premium principles]\label{PP} 
Insurance and reinsurance premiums are evaluated under a premium calculation principle. 
Here we recall a couple of celebrated formulas which are widely used in the actuarial field, namely the expected value principle (EVP) and the variance principle (VP). Other choices of the premium calculation principles are discussed, for instance, in \cite{Grandell} and \cite{Schmidli2018}.

We begin by observing that, in the Cram\'er-Lundberg model,  expectation and variance of the cumulative loss are calculated as $\mathbb{E}[C_t] = \mathbb{E}[Z_1] \lambda t$ and  $Var(C_t) = \mathbb{E}[Z^2_1] \lambda t$, respectively, where $Z_1$ is the representative amount of one claim.

Assume that both the insurance and the reinsurance premium are computed under the EVP, then they are given by 
\begin{align}
c & = (1 + \alpha^{(I)}) \mathbb{E}[Z_1] \lambda,\\
q(u) & = (1 + \alpha^{(R)}) (1-u)\mathbb{E}[Z_1] \lambda,
\end{align}
for any fixed $u\in [0,1]$, with  $\alpha^{(R)} > \alpha^{(I)}$ being the contractually decided reinsurance and insurance safety loadings, respectively. 

If premiums are evaluated under the VP, it holds that 
\begin{align}
c &= \mathbb{E}[Z_1] \lambda + \alpha^{(I)} \mathbb{E}[Z^2_1] \lambda,\\ q(u) &= (1-u)\mathbb{E}[Z_1] \lambda + \alpha^{(R)} (1-u)^2 \mathbb{E}[Z^2_1] \lambda, 
\end{align}
for every fixed $u \in [0,1]$. 
\end{remark}

\subsection{The financial market}
The insurer invests its surplus in a financial market which consists of a money market account and one risky asset\footnote{The extension to $N$ risky assets follows standard arguments and the results shown here carry over.}.
We let $r>0$ be the constant riskless interest rate, $\sigma_1>0$ be the volatility of the stock price and $X=\{X_t\}_{t \in [0,T]}$ be the market price of risk. 
The money market account has value process $B=\{B_t\}_{t \in [0,T]}$, and the price process of the risky asset is $S=\{S_t\}_{t \in [0,T]}$, which evolve according to the following system of SDEs
\begin{align}
\ud B_t & = r B_t \ud t, \quad B_0 =1\\
\label{priceS}\ud S_t &= S_t \Big ( \mu(X_t) \ud t + \sigma_1 \ud W^1_t \Big), \quad  S_0 \in \R^+,\end{align}
where $W^1$ is a Brownian motion and for every time $t\in [0,T]$
\begin{equation}\mu(X_t) = r + \sigma_1 X_t. \end{equation}

The market price of risk is assumed to follow a linear, mean reverting dynamics 
\begin{equation}\label{X}
\ud X_t= b_0(\mu_0 - X_t) \ud t + \sigma_0 \ud W^0_t,\end{equation}
where $W^0$ is a standard Brownian motion, correlated with $W^1$, with correlation coefficient $\rho\in[-1,1]$. The three parameters $b_0$, $\mu_0$ and $\sigma_0$ are strictly positive and represent, respectively the speed of mean reversion, the long run mean and the volatility of the process $X$. We also assume that $X_0$ is a Gaussian random variable with mean $\Pi_0\in \R$ and variance $P_0>0$.

\begin{remark} 
It is useful to note that the price process in equation \eqref{priceS} satisfies
$S_t = S_0 e^{Y_t}$, for every $t \in [0,T]$, with  $Y=\{Y_t\}_{t \in [0,T]}$ being the logreturn process,  
\begin{equation} \label{Y} Y_t = \int_0^t (\mu(X_s) - \frac{1}{2} \sigma^2_1 ) \ud s + \sigma_1 W^1_t, \quad t \in [0,T].\end{equation}
Therefore in the sequel we will indifferently work with $S$ or $Y$, depending which is more convenient.
\end{remark}

Under appropriate conditions on the model parameters, the aforementioned financial  market is free of arbitrage opportunities. Such conditions involve assuming that the process $X$ is relatively \emph{stable}, meaning that it does not deviate too much from the long run mean. This scenario is verified when the volatility remains adequately  small and the mean reversion speed large.

\begin{lemma}\label{Nov}
Under the assumption
\begin{equation}\label{nn}
P_0 + \frac{\sigma^2_0}{2 b_0} < \frac{1}{T},
\end{equation}
the Novikov condition 
\begin{equation}\label{Nov1}
\esp{e^{\frac{1}{2} \int_0^T X^2_t \ud t}} < \infty
\end{equation}
is satisfied.
\end{lemma}
The proof of this result is provided in Appendix \ref{app:technical}. 

The condition \eqref{Nov1} ensures the existence of an equivalent martingale measure for the financial market. The financial market is anyhow incomplete, since the randomness due to the stochastic market price of risk is not hedgeable. 

One equivalent martingale measure $\mathbf{Q}$ is characterised by the density $\left.\frac{\ud \mathbf{Q}}{\ud \P}\right|_{\mathcal{G}_T} = L_T$ where   
 \begin{equation} \label{mgm} 
  L_t=e^{-\frac{1}{2}\int_0^t X^2_s \ud s-\int_0^t X_s \ud W^1_s}, \quad t \in [0,T]
 \end{equation}
is a martingale under the filtration $\bG$ and the probability $\P$,  with expected value equal to $1$.

The dynamics of the risky asset price process can be equivalently expressed through the following differential equation: 
 $$\ud S_t = S_t ( r \ud t + \sigma_1 \ud W^{1,Q}_t ), \quad  S_0 \in \R^+$$
where the process $ W^{1,Q}_t = W^1_t + \int_0^t X_s \ \ud s$ is a Brownian motion under the measure $\mathbf{Q}$ and the filtration $\bG$. 
\subsection{The partial information framework}

We assume that the market price of risk is not directly observable by the insurer. Consequently the insurer  has no direct information on the return of the risky asset $\mu(X)$. 

The available information flow comes from the observation of two processes: the cumulative claims process $C$ and the risky asset price $S$. 

Therefore we introduce the \emph{observation filtration} $\bF$ as
\begin{equation}
\bF = \bF^C \vee \bF^S \subset \bG, 
\end{equation}
where, $\bF^C = \{\F_t^C\}_{t \in [0,T]}$ and $\bF^S = \{\F_t^S\}_{t \in [0,T]}$ are, respectively, the natural filtrations of the claim process and the asset price process. We assume that  $\bF$ is completed with $\P$-null sets and right continuous. We also observe that $\bF^S=\bF^Y$.

\subsection{The optimization problem}

We consider an investment strategy $\theta=\{\theta_t\}_{t \in [0,T]}$, where $\theta_t$ indicates the investment in the risky asset at time $t$, and let $a= \{a_t\}_{t \in [0,T]} =\{(u_t, \theta_t)\}_{t \in [0,T]}$ be the reinsurance and investment strategy. Technical conditions on the processes $u$ and $\theta$ are collected in the definition of admissible strategies.

The insurer's wealth associated with a given strategy $a =(u, \theta)$ follows the SDE
$$dZ^a_t   = dR^u_t + \theta_t \frac{\ud S_t}{S_t} + (Z^a_t - \theta_t)\frac{\ud B_t}{B_t},  \quad Z^a_0= R_0,$$ that is, explicitly,
\begin{align} 
dZ^a_t  & = \bigl(c-q(u_t)+ Z^a_t r + \theta_t  \sigma_1 X_t \bigr) \,dt  - u_t\int_{\R^+} \!z m(\ud t, \ud z) + \theta_t \sigma_1 \ud W^1_t \label{Za}
\end{align}
with the initial condition $Z^a_0= R_0$.
 
The insurer aims to maximize the expected exponential utility of terminal wealth, that is to solve 
\begin{equation}\label{exp_utility}
\sup_{a\in \mathcal{A}}\mathbb{E}\bigl[ 1-e^{-\eta Z^a_T} \bigr],
\end{equation}
where $\eta>0$ represents the parameter of risk aversion and  $\mathcal{A}$ indicates the class of admissible strategies as in Definition \ref{def:admissible}. 

\begin{definition}\label{def:admissible}
The set $\mathcal{A}$ consists of all processes $(u, \theta)$ where $u$ is $\bF$-predictable\footnote{Predictability is needed to get that stochastic integrals with respect to the jump measure $m(\ud t, \ud z)$ are well defined.} with values in $[0,1]$, $\theta$ is $\bF$-adapted with values in $\R$, and the conditions below hold:
\begin{equation}\label{admissible}
 \esp{\int_0^T  \theta^2_t \ud t} < \infty \ \mbox{ and } \ \sup_{t \in [0,T]}\mathbb{E}[e^{-2\eta (1+\epsilon) e^{r(T-t)} Z^a_t }] < \infty
 \end{equation}
 for some $\epsilon>0$.
 \end{definition}

An important characteristics of the optimization problem \eqref{exp_utility} is that the supremum is taken over a set of strategies that needs to be adapted to the observation filtration $\bF$, rather then to the global filtration $\bG$. 
Therefore we are facing a stochastic optimization problem in the context of partial information. To tackle this, we employ the methodology introduced by \cite{fleming1982optimal}, and introduce the auxiliary separated problem. Under the separated problem, all state variables are adapted to $\bF$. This step requires results from stochastic filtering theory, which are discussed in the next section.

\section{Reduction to the observable filtration}\label{sec:reduction}

The financial market is mathematically described in terms of a \emph{partially observable} pair of processes $(X, S)$ where $X$ is interpreted as an unobservable signal and $S$ is the observation process. As outlined before, we can equivalently replace $S$ with its logreturn $Y$, which has the advantage to be linear.

The conditional distribution of $X$, given the observation filtration is defined by $\esp{f(X_t) | \mathcal F_t}$ for every $t \in [0,T]$ and for every function $f$ such that $\esp{|f(X_t)|}< \infty$. In view of independence between the claim processes $C$ and the logreturn $Y$, the conditional expectation simplifies to $\esp{f(X_t)|\mathcal{F}^Y_t}$, and in view of the dynamics of $(X, Y)$ in equations \eqref{X} and \eqref{Y}, we fall in the setup of the celebrated Kalman-Bucy filter (\cite{bucy1967optimal}). The conditional distribution of $X$ is Gaussian, hence identified by its conditional expectation and the conditional variance, denoted respectively by $\Pi$ and $P$: 
\begin{equation}
\Pi_t=\esp{X_t|\F^Y_t} \mbox{ and } P_t= \esp{(X_t-\Pi_t)^2|\F^Y_t}, \quad t \in [0,T]. 
\end{equation}

Applying standard result of filtering theory (see, e.g. \cite[Theorem 10.3]{LipS}), we get that the pair $(\Pi, P)$ is the unique solution to the following system: 
\begin{align} 
\ud \Pi_t&= b_0(\mu_0 - \Pi_t) \ud t + \left(P_t +\rho \sigma_0 \right) \ud I_t \label{Kal}\\
\frac{\ud P_t}{\ud t}&=\sigma_0^2-2b_0P_t-\left(P_t+\rho \sigma_0\right)^2 \label{Ric}
\end{align}
with the initial conditions $\Pi_0$ and $P_0$ and where $I$ given by  
\begin{equation}\label{INN1}
I_t=W^1_t+\int_0^t (X_s-\Pi_s)\ud s, \quad t \in [0,T], 
\end{equation} 
is a Brownian motion under the filtration $\bF$ and the probability $\P$, called the innovation process.   
In particular, the process $\Pi$ is the unique $\bF$-adapted solution of a linear SDE; 
$P$ solves a deterministic Riccati equation, hence also satisfies $P_t= \esp{(X_t-\Pi_t)^2}$.

Using the definition of the innovation process we can equivalently write the dynamics of the asset price and the wealth as  
\begin{align}\label{eq:prices_partial}
\ud S_t &= S_t \Big( (r + \sigma_1 \Pi_t)  \ud t + \sigma_1 \ud I_t\Big ) , \quad  S_0 \in \R^+,\\
\label{Za2}
dZ^a_t &= \bigl(c-q(u_t) + Z^a_t r + \theta_t \sigma_1 \Pi_t \bigl)\,dt + \theta_t \sigma_1  \ud I_t - u_t\int_{\R^+} z m(\ud t, \ud z), \quad Z^a_0= R_0.
 \qquad{}
\end{align} 
Note that in these equations the unobservable market price of risk does not appear anymore, and all coefficients are adapted to the observation filtration $\bF$. 

It is also worth observing that the financial market under the partial information perspective, i.e. on the filtered probability space $(\Omega, \mathcal{F}_T, \P, \mathbb{F})$, is complete as the only source of randomness that drives the market is the innovation process. The unique equivalent martingale measure on the filtration is the restriction of the measure $\mathbf{Q}$ to $\mathcal{F}_T$, with the density process 
\begin{equation}\label{eq:Lhat}
\widehat L_t = e^{-\frac{1}{2}\int_0^t\Pi_s^2 \ud s - \int_0^t\Pi_s \ud I_s}, \quad t \in [0,T]. 
\end{equation}
Indeed, using the expression of the innovation process in equation \eqref{INN1} and Girsanov theorem we see that
\begin{equation}\label{eq:IQ}
\ud I^Q_t := \ud I_t + \Pi_t \ud t = \ud W^1_t+ X_t \ud t = \ud W^{1,Q}_t
\end{equation}
is a $(\mathbf{Q}, \bF)$-Brownian motion. Hence, it holds that \begin{equation}  \widehat{L}_t := \left. \frac{\ud \mathbf{Q}}{\ud \P}\right|_{\mathcal{F}_t} =  \mathbb{E}[ L_T | \F_t], \quad t \in [0,T].
 \end{equation}

Moreover for any constant strategy, $a=(u,\theta) \in [0,1] \times \R$, the couple $(Z^a, \Pi)$ is Markovian with respect to the filtration $\bF$ and the probability $\P$, and its infinitesimal generator is given below.

\begin{lemma}
 For any constant strategy, $a=(u,\theta) \in [0,1] \times \R$, the   
infinitesimal generator of $(Z^a, \Pi)$  is
\begin{align}
{\L}^a f(t,\zeta ,p)= & \frac{\partial f}{\partial t}(t, \zeta, p) + \bigl(c-q(u) + \theta \sigma_1 p + r \zeta  \bigl)\frac{\partial f}{\partial \zeta}(t, \zeta, p) + \frac{1}{2}\theta^2 \sigma_1^2 \frac{\partial^2 f}{\partial \zeta^2}(t, \zeta, p) + \\
&\int_{\R^+} [ f(t, \zeta -u z, p) - f(t,\zeta,p)] \lambda F_Z(\ud z) +  b_0(\mu_0 -p)\frac{\partial f}{\partial p}(t, \zeta, p) + \\
      &\frac{1}{2}\left(P_t +\rho \sigma_0 \right)^2 \frac{\partial^2 f}{\partial p^2}(t, \zeta, p) +
\left(P_t +\rho \sigma_0 \right) \theta \sigma_1 \frac{\partial^2 f}{\partial \zeta \partial p}(t, \zeta, p).
\end{align}
The domain $D^{\mathcal L}$ of the generator contains functions  $f \in C^{1,2,2}((0,T) \times \mathbb{R}^2)$ such that
\begin{align}
&\mathbb{E}\left[\int_0^{T} \left(\Big(\frac{\partial f}{\partial \zeta} (s,Z^a_s, \Pi_s) \Big)^2 + \Big(\frac{\partial f}{\partial p} (s,Z^a_s, \Pi_s) \Big)^2 \right)\ud s\right]<\infty\\
& \mathbb{E}\left[\int_0^{T} \int_{\R^+} \left|f(s, Z^a_{s-} - u z, \Pi_s) - f(s, Z^a_{s-}, \Pi_s) \right|  F_Z(\ud z) \ud s\right]< \infty
\end{align}
\end{lemma}

The proof is an immediate application of It\^o's formula. 

In the sequel we use the shorter notation $\mathbb{E}_{t,\zeta,p}[ \cdot ]$ for the conditional expectation, given the initial values of the state variables, $Z^a_t=\zeta$ and $\Pi_t=p$, with the starting time $t$ and introduce the separated problem 
$$\sup_{a\in \mathcal{A}}\mathbb{E}_{t,\zeta,p}\bigl[ 1-e^{-\eta Z^a_T} \bigr] = 1 - \inf_{a\in \mathcal{A}}\mathbb{E}_{t,\zeta,p}\bigl[ e^{-\eta Z^a_T} \bigr].$$

We denote by $v(t,\zeta,p)$ the value function 
\begin{equation}v(t,\zeta,p) := \inf_{a\in \mathcal{A}}\mathbb{E}_{t,\zeta,p}\bigl[ e^{-\eta Z^a_T} \bigr], \label{pb:optimization}
\end{equation}
which satisfies  
\begin{equation} \label{eq:separated_pb}
\sup_{a\in \mathcal{A}}\mathbb{E}\bigl[ 1-e^{-\eta Z^a_T} \bigr] = 1 - v(0,R_0, \Pi_0).
\end{equation}
Our goal in the next section is to characterize the function $v(t, \zeta, p)$ as well as the optimal strategy $a^*\in \mathcal{A}$ at which the supremum in \eqref{eq:separated_pb} is achieved. 

\section{The Hamilton-Jacobi-Bellman approach}\label{sec:HJB}
Suppose for the time being, that the function $v(t,\zeta,p)$ is differentiable with respect to $t$ and twice differentiable with respect to $\zeta$ and $p$. Then $v(t,\zeta, p)$ must be the solution of the HJB equation 

\begin{equation}\label{HJB}
 \inf_{a \in [0,1] \times \mathbb{R}} {\L}^a v(t,\zeta,p) =0
\end{equation}
with the final condition $v(T,\zeta,p) = e^{-\eta \zeta}$.

Conversely, the following  verification result holds.  

\begin{theorem}[Verification theorem]\label{thm:verification}

Let $f(t,\zeta, p) \in C^{1,2,2}([0,T] \times \mathbb{R}^2)$ be a classical solution to 
the HJB \eqref{HJB} and assume the following conditions:

\begin{itemize}
\item[(i)] for any $a \in \mathcal{A}$ the family $\{f(\tau\wedge T, Z^a_{\tau\wedge T}, \Pi_{\tau\wedge T}): \tau \mbox{ is an } \bF-\mbox{stopping time}\}$ is uniformly integrable; 
\item[(ii)] there exists  $a^*(t,p,\zeta) = (u^*(t,p,\zeta), \theta^*(t,p,\zeta))$ at which  the infimum in equation \eqref{HJB} is attained. 
\end{itemize}

Then $f(t,\zeta, p)=v(t,\zeta, p)$ and if  $\{a^*(t,Z^{a^*}_{t^-},\Pi_t)\}_{t \in [0,T]} \in \mathcal{A}$, this is an optimal Markovian control.
\end{theorem}
The proof of this result is given in Appendix \ref{app:technical}.

In view of the Verification theorem, we will characterize the value function as a solution to the HJB equation. Due to the characteristics of the problem we guess that $v(t, \zeta, p)$ has the form 
\begin{equation}\label{Ansatz}v(t,\zeta,p) = e^{-\eta \zeta e^{r(T-t)}} h(t) e ^{\psi(t,p)}\end{equation}
for suitable functions $h(t)$ and $\psi(t,p)$.

If \eqref{Ansatz} holds, then equation \eqref{HJB} splits into an ordinary differential equation and a partial differential equation given, respectively by equations \eqref{Rein} and \eqref{Inv1} below:  
\begin{equation}\label{Rein}
h'(t) + h(t) \inf_{u \in [0,1]} H^R(t,u) =0,
\end{equation}
with the final condition $h(T)=1$ and 
\begin{equation}\label{Inv1}
\frac{\partial \psi}{\partial t} (t, p) + b_0 (\mu_0-p) \frac{\partial \psi}{ \partial  p} (t, p)  + \frac{1}{2}\left(P_t +\rho \sigma_0\right)^2 \Big [ \frac{\partial^2 \psi }{ \partial p^2} (t, p) + 
\Big (\frac{\partial \psi }{\partial  p} (t, p) \Big )^2  \Big ]
+ \inf_{\theta \in \R} H^I (t, \theta, p) =0 
\end{equation}
with the final condition $\psi(T,p)=0$, where we have put
\begin{align*}
H^R(t, u)&:=(c-q(u)) \eta e^{r(T-t)} + \int_{\R^+}  \bigl( e^{\eta u z e^{r(T-t)}} -1\bigr) \lambda F_Z(\ud z),\\
H^I (t, \theta, p)  &:= {\frac{1}{2}}\theta^2 \big (\sigma_1 \eta e^{r(T-t)} \big )^2 - \eta e^{r(T-t)} \theta \sigma_1 \Big (p  + 
\left(P_t +\rho \sigma_0 \right) \frac{\partial \psi }{\partial p}(t, p) \Big ).
\end{align*}

Equation \eqref{Ansatz} allows to separate the optimization problem in two sub-problems with a key feature: the first optimization involves only reinsurance, the second optimization only investment. Such simplification is a mere consequence of independence of the financial and the actuarial setups. 
It is also wort observing that in the first optimization the control  may take values in a compact set, and hence the existence of the optimizer is guarantees by the fact that the function $H^R$ is continuous in $u \in [0,1]$. In the second optimization the existence of the optimizer, instead, is guaranteed by convexity of $H^I$ in $\theta \in \R$. 

In the next two section we address the optimal reinsurance and the optimal investment problems separately.

\subsection{Optimal Reinsurance}
We begin with a discussion on the optimization problem  \eqref{Rein}. 

Assuming that the reinsurance premium $q(u)$ is continuous for all $u \in[0,1]$ guarantees that a minimiser of $H^R(t,u)$ exists. We denote the minimiser by $\{u^*(t)\}_{t \in [0,T]}$, then it holds that 
\begin{equation}\label{h}
h(t) = e^{\int_t^T H^R(s, u^*(s)) \ud s}.
\end{equation} 

\begin{remark}[A simple example with the expected value premium calculation principle.]
In case the reinsurance premium is calculated under the expected value principles recalled in Remark \ref{PP}, the function $ H^R(t,u)$ is strictly convex for all $u \in[0,1]$, and then a unique minimizer is given by 
\begin{equation}\label{u*}
u^*(t) =
\begin{cases}
\bar{u}(t)			& \alpha^{(R)} < \bar\alpha(t) \\
1			& \alpha^{(R)} \geq \bar\alpha(t),
\end{cases}
\end{equation}
with $\bar\alpha(t) = \frac{1}{\mathbb{E}[Z_1]} \int_{\R^+}  z e^{\eta z e^{r(T-t)}}F_Z(\ud z) -1$, 
and where $\bar u(t)$  being the unique solution of the first order condition
$$(1 + \alpha^{(R)}) \mathbb{E}[Z_1] = \int_{\R^+}  z e^{\eta u z e^{r(T-t)}}F_Z(\ud z) = 
\mathbb{E}[Z_1 e^{\eta u Z_1 e^{r(T-t)}}].$$ 

The control $u^*(t)$ is the candidate optimal reinsurance (precisely, it becomes the optimal retention level if Verification theorem applies). Under this premium calculation principle it has the following interpretation: if reinsurance contract is too expensive, namely the safety loading $\alpha^{(R)}$   exceeds a threshold function $\bar\alpha(t)$, it is not convenient to buy reinsurance, and hence,  null reinsurance is optimal; otherwise, if the reinsurance contract is cheap enough,  the optimal retention level will be $0<u^*(t)<1$, that is, the ceding company finds convenient to transfers to a reinsurer a percentage $1-u^*(t)$ of risk. 
\end{remark}

\subsection{Optimal Investment}
In this section  we consider the investment  problem  \eqref{Inv1}. 
Taking the first and second order conditions with respect to $\theta$ on the function $H^I (t, \theta, p)$ we get that   

\begin{equation}\label{theta*}\theta^*(t,p) = \frac{ p }{\sigma_1 \eta} e^{-r(T-t)} + \frac{P_t + \rho \sigma_0  }{\sigma_1 \eta} e^{-r(T-t)} \frac{\partial \psi }{\partial p}(t, p). 
\end{equation}

Replacing $\theta$ with $\theta^*$ in equation \eqref{Inv1} leads to 

\begin{equation}\label{Inv2}
\frac{\partial \psi }{\partial t} (t, p) + \Big [b_0(\mu_0 -p) - p(P_t +\rho \sigma_0) \Big ] \frac{\partial \psi }{\partial  p} (t, p)  + \frac{1}{2}\left(P_t +\rho \sigma_0\right)^2 \frac{\partial^2 \psi }{\partial p^2} (t, p) - 
\frac{1}{2}p^2 =0.
\end{equation}
with the final condition $\psi(T,p)=0$. Equation \eqref{Inv2} has a unique classical solution, with the explicit stochastic representation computed by applying Feynman-Kac formula (see, for instance, \cite[Theorem 1.3.17]{pham}). Letting $\widetilde \Pi^{t,p}= \{\widetilde \Pi^{t,p}_s, s \in [t,T]\}$ be the solution of the SDE starting from $p$ at time $t$ given by
\begin{equation}\label{FK}
\ud \widetilde \Pi^{t,p}_s= (b_0\mu_0 - a(s) \widetilde \Pi^{t,p}_s)\ud s + \left(P_s +\rho \sigma_0 \right) \ud I_s, \quad \widetilde \Pi_t^{t,p}=p
\end{equation}
where $a(t) := b_0 + P_t + \rho \sigma_0$ for all $t \in [0,T]$, then  the Feynman-Kac representation of the solution of equation \eqref{Inv2} is  
\begin{equation}\label{eq:psi}
\psi(t,p) = -{\frac{1}{2}}\mathbb{E} \left[\int_t^T \widetilde \Pi^{t,p}_s \ud s \right],
\end{equation}
for all $(t,p) \in [0,T]\times \R$. Next we apply It\^o's formula to the process $\widetilde \Pi_s^{t,p} e^{\int_t^s a(l) \ud l}$, which implies that the solution of equation \eqref{FK}, for all $s>t$, is 
$$\widetilde \Pi^{t,p}_s = p e^{-\int_t^s a(l) \ud l} + b_0\mu_0 \int_t^s e^{-\int_v^s   a(l) \ud l} \ud v +
\int_t^s \left(P_v +\rho \sigma_0 \right) e^{- \int_v^s a(l) \ud l} \ud I_v.$$
Taking expectation 
$$\mathbb{E}[ \widetilde \Pi^{t,p}_s ] = p e^{-\int_t^s a(l) \ud l} + b_0\mu_0 \int_t^s e^{-\int_v^s a(l) \ud l} \ud v,$$
and substituting in  \eqref{eq:psi} we get
\begin{equation} \label{PSI}\psi(t,p) = -{\frac{1}{2}} \int_t^T \Big \{ p e^{-\int_t^s a(l) \ud l} + b_0\mu_0 \int_t^s e^{-\int_v^s a(l) \ud l} \ud v \Big \} \ud s.
\end{equation}
This also confirms that $\psi(t,p) \in C^{1,2}([0,T] \times \mathbb{R}).$

\begin{remark}\label{rapp}
We observe that the law of the process $\widetilde \Pi^{t,p}$ under the probability measure $\P$ coincides with the law of the process $\Pi^{t,p}$ under the martingale measure $\mathbf{Q}$ with Radon-Nykodim density given in equation \eqref{mgm}. This follows from the fact that the process $I^\Q$ in equation \eqref{eq:IQ} is a $(\mathbf{Q}, \bF)$-Brownian motion and that $\Pi^{t,p}$ solves the equation 
$$
\ud \Pi^{t,p}_s= (b_0\mu_0 - a(s)  \Pi^{t,p}_s) \ud s + \left(P_s +\rho \sigma_0 \right) \ud I^Q_s, \quad \Pi^{t,p}_t =p. $$
\end{remark} 

\subsection{Solution to the optimal  reinsurance and investment problem} 
In view of the Verification Theorem \ref{thm:verification}, the candidate as the value function is 
\begin{equation}\label{v1}
v(t,\zeta,p)= e^{-\eta \zeta e^{r(T-t)}} h(t) e ^{\psi(t,p)}\end{equation} 
with the functions $ h(t)$ and $\psi(t,p)$ given, respectively, in equations \eqref{h} and \eqref{PSI}.
Moreover, the candidate optimal strategy is $a^*=(u^*, \theta^*)$. The first component 
 $u^*_t=u^*(t)$ is deterministic, and for example, under the expected value premium principle is given by \eqref{u*}. The second component $\theta^*_t = \theta^*(t, \Pi_t)$ is Markovian, and depends on the filtered estimate of the market price of risk. Explicitly we have that  
\begin{equation}\label{tetapartial} \theta^*_t =  \frac{ \Pi_t }{ \sigma_1 \eta} e^{-r(T-t)} - {\frac{1}{2}} \frac{P_t + \rho \sigma_0  }{\sigma_1 \eta} e^{-r(T-t)} \int_t^T e^{-\int_t^s (b_0 + P_l + \rho \sigma_0) \ud l} \ud s, \end{equation}
for every $t \in [0,T]$. 

It remains to show that the Verification theorem applies and that this candidate optimal strategy is admissible. These problems are addressed in the reminder of the paper.  

\section{Verification of admissibility of the optimal strategy}\label{sec:verification}

In this section we want to give conditions on the parameters of the financial-insurance market model which guarantee that condition (i) of Theorem \ref{thm:verification} is satisfied and that  $a^*=(u^*, \theta^*)$ is an admissible control, according to Definition \ref{def:admissible}. 

Our fist lemma is a preliminary result. The proof is given in Appendix \ref{app:technical}. 

\begin{lemma}\label{squareMGM}
Assume that $\mathbb{E}^{\Q}\Bigl[e^{6\int_0^T \Pi^2_t \ud t} \Bigr]< \infty$. 
Recall that the process $\widehat L$ defined in \eqref{eq:Lhat} is the density of the measure $\Q$ on $\bF$   
\begin{equation}  
 \widehat{L}_t := \left. \frac{\ud \mathbf{Q}}{\ud \P}\right|_{\mathcal{F}_t} =  \mathbb{E}[ L_T | \F_t], \quad t \in [0,T].
 \end{equation}
Then, it holds that  
$$\mathbb{E}^{\Q}[\widehat L^{-2}_T] < \infty.$$

\end{lemma}

Next we will use this result to show that condition (i) in the Verification theorem applies. 

\begin{proposition}\label{uniform}
Suppose that $\mathbb{E}^{\Q}\Bigl[e^{\int_0^T 6 \Pi^2_t \ud t} \Bigr]< \infty$. 
Let $v(t,\zeta,p)$ be the function given in equation \eqref{v1}. Then,  for any admissible strategy $a \in \A$,  $$\{ v(\tau, Z^a_{\tau}, \Pi_{\tau}) ; \ \mbox{for all } \  \bF-\mbox{stopping times } \ \tau\le T\}$$ forms a uniformly integrable family. 
\end{proposition}

In the next proposition we provide sufficient conditions for admissibility of the optimal strategy. Proof is postponed to Appendix \ref{app:technical}.

 \begin{proposition}\label{adm_star}
We fix $K=16 (1+\epsilon)^2 e^{2rT}$ for some $\epsilon>0$ and assume that $\mathbb{E}^{\Q}\Bigl[e^{K \int_0^T \Pi^2_t \ud t} \Bigr]< \infty$. Then the process  $a^*=(u^*, \theta^*)$ is an admissible strategy. 
\end{proposition}

We conclude this section with some sufficient conditions on the parameters of the model aimed to guarantee that Proposition  \ref{adm_star} and Proposition  \ref{uniform} apply. 

\begin{proposition}\label{prop:adm2}
Let $\bar P = \sup_{t \in [0,T]} P_t$ and $K$ as in Proposition \ref{adm_star}. Suppose
\begin{equation}\label{nn1}
 \frac{(\bar P + \rho\sigma_0)^2}{ b_0 + \rho\sigma_0} < \frac{1}{T K}.
\end{equation}
Then $\mathbb{E}^{\Q}\Bigl[e^{K \int_0^T \Pi^2_t \ud t} \Bigr]< \infty$.
\end{proposition}

To summarise, under the assumption of Proposition \ref{prop:adm2}, the candidate optimal strategy is admissible and the value function is the unique solution of the optimization problem. It is important to note that, although this assumption may seem to be restrictive, it offers the significant advantage of being explicit, as it depends only on the model parameters. Therefore it can be easily verified, which is also consistent with the approach of examining a simple model.

\section{Comparison results}\label{sec:comparison}
We would like to compare in this section the optimal strategies followed by a fully informed insurer and by a partially informed insurer. 

In our numerical experiments we fix the parameters according to the values in Table \ref{Tab:parameters}, unless otherwise stated.

\begin{table}[ht]
\begin{tabular}{|l|l|l|l|l|l|l|l|l|}
\hline
$b_0$ & $\mu_0$ & $\sigma_0$ & $\sigma_1$ & $\Pi_0$ & $P_0$ & $T$  & $r$ & $\eta$ \\ \hline
$1$   & $0.4$   & $0.18$      & $0.2$      & $0.4$   & $0.03$ & $10$ & $0$ & $0.5$  \\ \hline
\end{tabular}
\vspace{.3cm}
\caption{General parameters for the numerical study}\label{Tab:parameters}
\end{table}

\begin{figure}[ht]
    \centering
    \includegraphics[width=.9\textwidth]{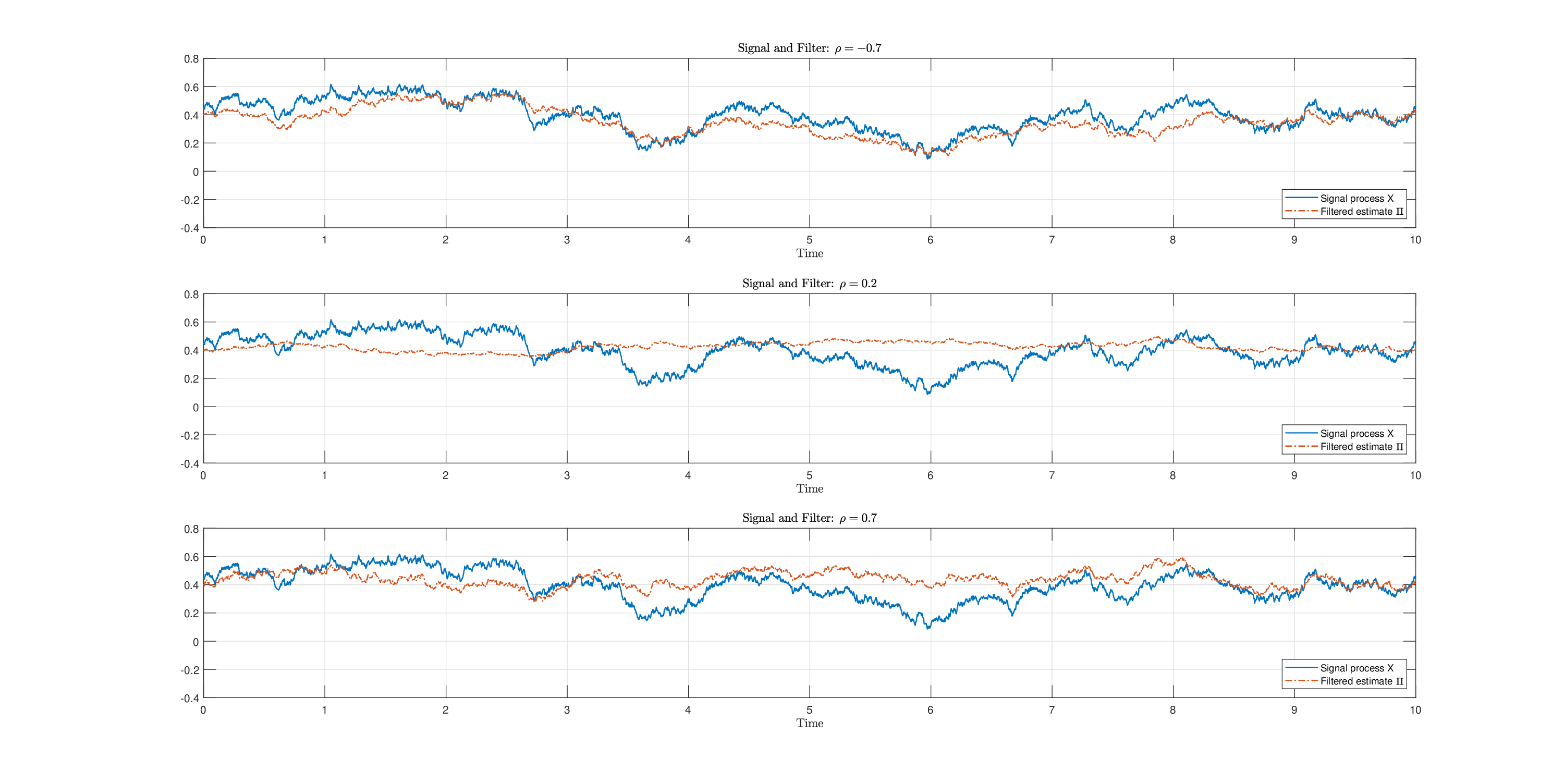}
    \caption{True trajectory of the market price of risk $X$, solid blue line, and trajectories of its filtered estimate $\Pi$, dashed red line, for different correlation values.}
    \label{fig:filtro}
\end{figure}

We begin with a representation of the comparison between the true trajectory of the market price of risk $X$ and its filtered estimate $\Pi$, which is shown in Figure \ref{fig:filtro}. We see that the estimation improves as the observation is more informative, i.e. for large absolute values of the correlation coefficient $\rho$. There is also some learning effect over time, that is better displayed for the upper panel and the lower panel, i.e. for larger correlations.

Next we would like to focus on strategies.
If the insurer has complete knowledge on the Market price of risk the optimal investment strategy can be derived by \eqref{tetapartial} (taking $X=\Pi$ and $P=0$)
\begin{equation}\label{tetafull} \theta^{*,F}_t =  \frac{ X_t }{\sigma_1 \eta} e^{-r(T-t)} - {\frac{1}{2}} \frac{\rho \sigma_0  e^{-r(T-t)}}{\sigma_1 \eta  (b_0 + \rho \sigma_0)} (1 -  e^{- (b_0 + \rho \sigma_0)(T-t)}),\end{equation}
which consists of two parts, namely the myopic component, corresponding to the first term,  and the hedging demand for parameter risk, corresponding to the second term, which depends on model parameters and the correlation coefficient $\rho$ between $X$ and the logreturn price $Y$. The hedging demand is clearly zero if the market price of risk and the price process are uncorrelated.

\begin{figure}[ht]
    \centering
\includegraphics[width=.9\textwidth]{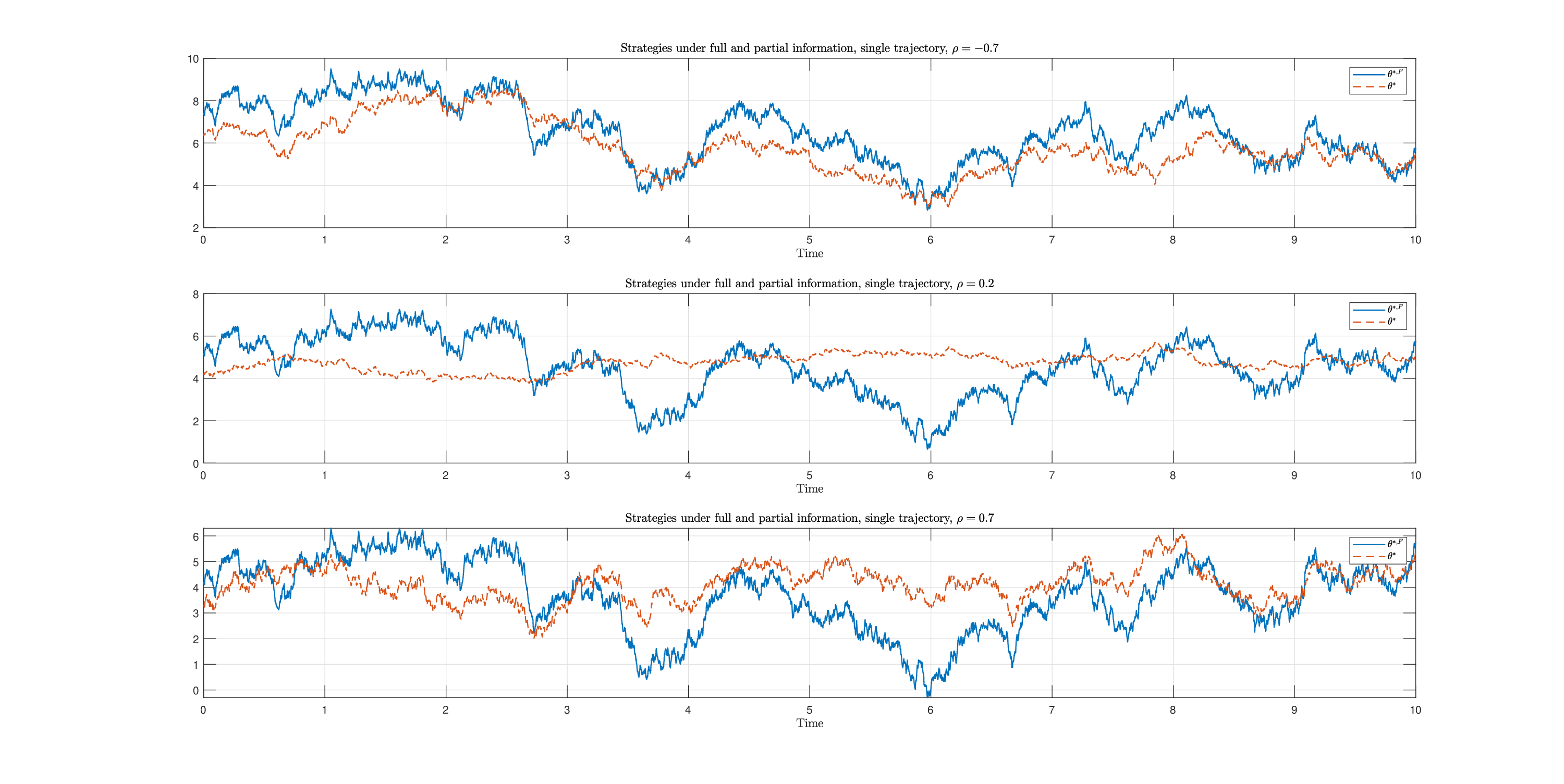}
    \caption{Trajectories of strategies under full and partial information for different values of the correlation.}
    \label{fig:strategies}
\end{figure}

If the insurer has only a limited knowledge on the market price of risk, she/he will follow the strategy described by equation \eqref{tetapartial}. Here we still recognize a myopic component, and the hedging demand with slightly different characteristics. In the myopic component the market price of risk is replaced by its filtered estimate, as expected; however, the structure of the hedging demand is more involved. It depends on the model parameters in a slightly different way, and most importantly, even if the dynamics of stock price and that market price of risk are uncorrelated, it does not vanish. This is a consequence of the fact that the price process is used to estimate the market price of risk. Mathematically, the dynamics of the price process and the filter are always perfectly correlated as they are both driven by the innovation process $I$, and the goodness of the estimate depends, roughly speaking, on how \emph{similar} $X$ and $S$ are (i.e. on $\rho$). For instance, in the cases $\rho=1, -1$ we get the \emph{best} estimates and the \emph{worst case}, instead, would be when $\rho=0$, see Figure \ref{fig:filtro}, and it therefore translates into strategies, as represented in Figure \ref{fig:strategies}.     
Recall also that $P_t = \mathbb{E}[ (X_t - \Pi_t)^2]$, hence this is the (deterministic) process that accounts for the mean square error between $X$ and its filtered estimate  $\Pi$.

It is easy to show that when $\rho=0$ the total amount invested in the risky asset at any time $t \in [0,T]$ in the partial information case, in expectation is smaller than the expected amount invested in the risky asset in the full information case, that is  $\mathbb{E}[\theta^{*}_t] < \mathbb{E}[\theta^{*,F}_t]$. Indeed, since for any $t\in [0,T]$ $\mathbb{E}[X_t] = \mathbb{E}[\Pi_t]$ we have that
$$\mathbb{E}[\theta^{*}_t] = \mathbb{E}[\theta^{*,F}_t]  - {\frac{1}{2}} \frac{P_t   }{\sigma_1 \eta} e^{-r(T-t)} \int_t^T e^{-\int_t^s (b_0 + P_l) \ud l} \ud s.$$
Finally, we also see that for any $t \in [0,T]$, and still in the case $\rho=0$, 
$$\theta^{*}_t =  \frac{ \Pi_t }{\sigma_1 \eta} e^{-r(T-t)}  - {\frac{1}{2}} \frac{P_t   }{\sigma_1 \eta} e^{-r(T-t)} \int_t^T e^{-\int_t^s (b_0 + P_l) \ud l} < \frac{ \Pi_t }{\sigma_1 \eta} e^{-r(T-t)} = \mathbb{E}[\theta^{*,F}_t | \F_t], \quad \P-a.s.$$

\begin{figure}
    \centering
    \includegraphics[width=.7\textwidth]{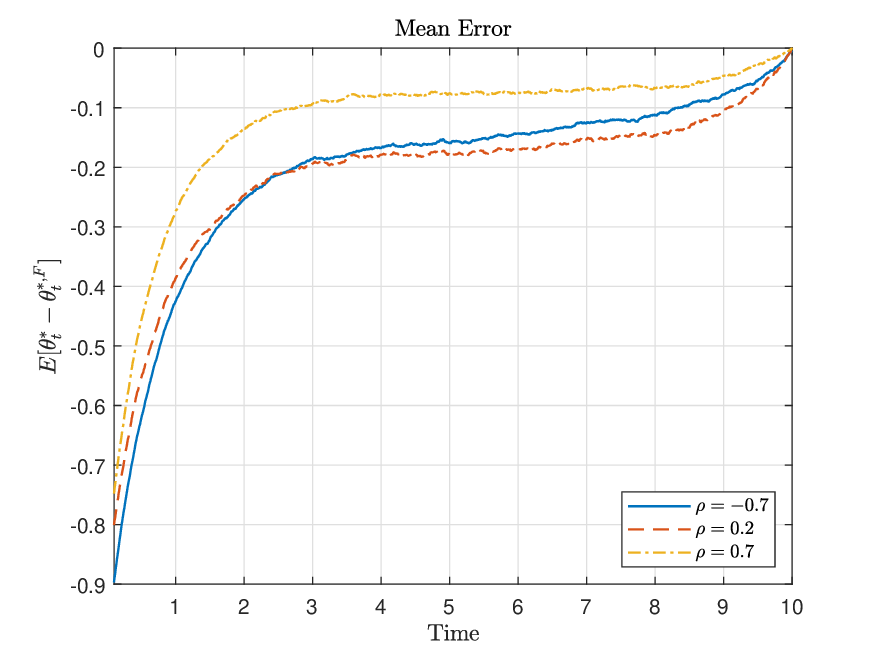}
    \caption{Mean error between strategies under full and partial information, for different value of the correlation}
    \label{fig:error}
\end{figure}

In Figure \ref{fig:error} we display the mean error between the strategy under partial information and full information, $\mathbb{E}[\theta^{*}_t - \theta^{*,F}_t]$, for several values of the correlation.  

The takeout message is then that in a general setup, the optimal investment strategy under full information is not obtained by replacing the market price of risk with its filtered estimate, but there is an additional estimation error depending on the conditional variance of the filter, that must be taken into account, and which provides and additional component on the hedging demand.  

We conclude this section by computing the so called \emph{indifference value of information}. This is the amount $\Delta \zeta$ that makes the investor indifferent between implementing the strategy under partial information and buying the possibility to implement the strategy under full information. This  criterion enables a comparison between the full and the partial information setup in a market coherent way. \footnote{Another interesting criterion is to evaluate the loss of utility which has been used, e.g. in  \citet{lee2016pairs} (see also \citet{altay2018pairs, altay2019optimal}, for the case where the hidden stochastic factor is a modeled as a Markov chain), to highlight, for instance, that poor estimate of the filter generates lower expected utility.} Precisely we define the indifference value as the solution of the equation 
\begin{equation}
    \sup_{a\in \mathcal{A}}\mathbb{E}\bigl[ 1-e^{-\eta Z^a_T(R_0)}\bigr]=\sup_{a\in \mathcal{A}^{full}}\mathbb{E}\bigl[ 1-e^{-\eta Z^a_T(R_0-\Delta\zeta)} \bigr] 
\end{equation}
where $Z^a_T(\zeta)$ indicates the final wealth generated by the strategy $a$ with initial endowment $\zeta$. 
Solving this problem is equivalent to 
\begin{equation}
\mathbb{E}\left[v(0,R_0, \Pi_0)\right]= \mathbb{E}\left[v^{full}(0, R_0-\Delta \zeta, X_0)\right]
\end{equation}
where $v^{full}(t, \zeta, x)$ is the solution of the optimization 
$\inf_{a\in \mathcal{A}^{full}}\mathbb{E}_{t,\zeta,x}\bigl[ e^{-\eta Z^a_T} \bigr]$, and $\mathcal{A}^{full}$ is the set of all processes $(u, \theta)$ where $u$ is $\bG$-predictable with values in $[0,1]$, $\theta$ is $\bG$-adapted with values in $\R$, and satisfy condition \eqref{admissible}. Note that the term on the left coincides with $v(0, R_0, \Pi_0)$. 
By replicating the same arguments as in Sections \ref{sec:HJB} and \ref{sec:verification}, we get that $u^*$ is also the optimal reinsurance strategy under full information, $\theta^{*,F}$ in equation \eqref{tetafull} is the optimal investment strategy under full information and that the value function has the form 
$v^{full}(t,\zeta,x)=e^{-\eta \zeta e^{r(T-t)}} h(t) e^{\phi(t,x)}$ where $\phi$ is given by 
\begin{equation}
    \phi(t,x)=-\frac{b_0\mu_0}{2(b_0+\rho \sigma_0)}(T-t)- \left(x-\frac{b_0\mu_0}{b_0+\rho \sigma_0}\right)\frac{1-e^{-(T-t)(b_0+\rho\sigma_0)}}{2(b_0+\rho\sigma_0)}. 
\end{equation}
Using explicit representations for the value functions of the problems under full and partial information we get that 
\begin{equation}
    \Delta \zeta = \frac{e^{-rT}}{\eta} \left(\psi(0, \Pi_0)-\ln\left(\mathbb{E}\left[e^{\phi(0, X_0)}\right]\right)\right), 
\end{equation}
where $\psi(t,p)$ is given in equation \eqref{PSI}. 
Since $X_0\sim N(\Pi_0, P_0)$, denoting $C_0=\frac{1-e^{-(b_0+\rho\sigma_0) T}}{2(b_0+\rho\sigma_0)}$, it is easy to see that 
\begin{equation}
    \Delta \zeta = \frac{e^{-rT}}{\eta} \left(\psi(0, \Pi_0)-\phi(0, \Pi_0) - \frac{P_0 C_0^2}{2}\right). 
\end{equation}
In particular it holds that, if the market price of risk is known at time $t=0$, i.e. $X_0=\Pi_0$, then the expression above simplifies into  
\begin{equation}\label{eq:deltazero}
    \Delta \zeta = \frac{e^{-rT}}{\eta} \left(\psi(0, \Pi_0)-\phi(0, \Pi_0)\right). 
\end{equation}
This can be interpreted as the value of a complete knowledge of the market price of risk over the time interval $[0,T]$. Table \ref{Tab:tab1} and Table \ref{Tab:tab2} illustrate the indifferent value of information for different initial uncertainty on the market price of risk, $P_0$, different correlations between the observation process and the market price of risk, $\rho$, and different maturities. Note that the initial uncertainty $P_0$, intervenes explicitly in the expression \eqref{eq:deltazero}, but it also affects the value od $\psi$ through the dynamics of the conditional variance.  We see that the value of information is increasing in the maturity. Indeed, intuitively, an investor would pay more to get information on the trajectory of the market price of risk for a longer period. Finally, Table \ref{Tab:tab2} confirms that low correlation leads to more expensive value of information, as this corresponds to the least informative case.

\begin{table}[ht]
\begin{tabular}{|l|l|l|l|}
\hline
         & $T=3$ & $T=5$ & $T=10$ \\ \hline
$P_0=0$   &  0.6336   &  1.5384   &  4.0055    \\ \hline
$P_0=0.01$ &  0.6362   & 1.5406    &   4.0194   \\ \hline
$P_0=0.03$ &  0.6411   & 1.5520    &   4.0469   \\ \hline
\end{tabular}
\vspace{.2cm}
\caption{Indifference value of information, $\Delta \zeta$, for different initial uncertainty and maturities}\label{Tab:tab1}
\end{table}

\begin{table}[ht]
\begin{tabular}{|l|l|l|l|}
\hline
         &$T=3$ & $T=5$ & $T=10$ \\ \hline
$\rho=0$   &  0.6336  &  1.5348  &  4.0055    \\ \hline
$\rho=0.2$  & 0.6308   & 1.5103    &   3.8999   \\ \hline
$\rho=0.7$  & 0.6206   & 1.4487    &   3.6589   \\ \hline
\end{tabular}
\vspace{.2cm}
\caption{Indifference value of information, $\Delta \zeta$, for different correlations and maturities}\label{Tab:tab2}
\end{table}

\section*{Acknowledgements}  
The authors are grateful to two anonymous referees for their suggestions and valuable comments that allowed to improve the quality of the manuscript. The work of the authors has been funded by the European Union - Next Generation EU - Project PRIN 2022 (code BEMMLZ) with the title \emph{Stochastic control and games and the role of information}. The authors are also members of the GNAMPA group within the Italian National Institute for high Mathematics (INdAM). 

\appendix

\section{Proofs}\label{app:technical}
\subsection{Proof of Lemma \ref{Nov}}
By Jensen's inequality $e^{\frac{1}{2} \int_0^T X^2_t \ud t} \leq\frac{1}{T} \int_0^T e^{\frac{T}{2}  X^2_t} \ud t$, therefore
$$\esp{e^{\frac{1}{2} \int_0^T X^2_t \ud t}} \leq \frac{1}{T}\int_0^T \esp{e^{\frac{T}{2} X^2_t}} \ud t.$$
Recall that $X$ is an Ornstein-Ulhenback process, with Gaussian initial condition. Hence at any time $t$, $X_t$ is Gaussian with mean $m(t) = \Pi_0 e^{-b_0 t} + {\mu_0}(1- e^{-b_0t})$ and variance $\sigma^2_X(t) = P_0 e^{-2 b_0 t} + \frac{\sigma^2_0}{2 b_0}(1- e^{-2 b_0t}),$
which are finite in $[0,T]$ since $b_0>0$. Hence, $\esp{e^{\frac{T}{2} X^2_t}}$ is finite if 
$\esp{e^{\frac{T}{2} \sigma^2_X(t) \chi^2} } < \infty.$
Recalling that the moment generating function of the $\chi^2$ distribution $\esp{e^{\theta \chi^2} }$ is finite for $\theta < \frac{1}{2}$, we conclude that \eqref{Nov1} holds  under condition \eqref{nn}.

\subsection{Proof of Theorem \ref{thm:verification}}

From It\^o's formula applied to $f(t, Z^a_t, \Pi_t)$ we get that, for any $0\leq t\leq T$  and $a =(u, \theta) \in \mathcal{A}$, it holds

\begin{equation}\label{eq:function_f}
    f(T, Z^a_T, \Pi_T) = f(t, Z^a_t, \Pi_t) + \int_t^T{\L}^a f(s,Z^a_s, \Pi_s) \ud s + M_T - M_t,
\end{equation}
where
\begin{align}
M_t = & \int_0^t \frac{\partial f}{\partial \zeta} (s,Z^a_s, \Pi_s) \theta_s \sigma_1 \ud I_s + \int_0^t \frac{\partial f}{\partial p} (s,Z^a_s, \Pi_s) (\rho \sigma_0 + P_s) \ud I_s + \\
& \int_0^t \int_{\R^+}  \bigl(f(s, Z^a_{s-} - u_sz, \Pi_s) - f(s, Z^a_{s-}, \Pi_s) \bigr) (m(\ud s, \ud z) - \lambda F_Z(\ud z)\ud s), \quad t \in [0,T].
\end{align}

Next we introduce the sequence of stopping times defined as 
\begin{equation}
\tau_n = \inf\{t\in [0,T] : |Z^a_t| \geq n \mbox{ or } |\Pi_t|\geq n \}. 
\end{equation}
These are $\mathbb{F}$-stopping times since both $Z^a_t$ and $\Pi_t$ are $\mathcal F_t$-measurable for all $t \in [0,T]$ and it holds that, for $n\to \infty$, $\tau_n\to T$ because the processes do not explode. 

By assumption, $f$ is a smooth solution of the HJB equation, which means that $f$, $\frac{\partial f}{\partial \zeta}$ and $\frac{\partial f}{\partial p}$ are continuous, hence bounded on compact sets. Therefore the stopped  process $\{M_{t\wedge \tau_n}\}_{t \in [0,T]}$ is a martingale under $\P$ and with respect to the filtration $\mathbb{F}$. 
Indeed, for every $n \in \bN$ the following conditions are satisfied
\begin{align}
&\mathbb{E}\left[\int_0^{\tau_n}\left(\frac{\partial f}{\partial \zeta} (s,Z^a_s, \Pi_s) \theta_s \sigma_1\right)^2 \ud s\right]\leq \sup_{t \le T, \ (\zeta,p) \in [-n,n]^2}\left|\frac{\partial f}{\partial \zeta} (t, \zeta, p)\right|^2 \mathbb{E}\left[\int_0^T\sigma^2_1\theta_s^2 \ud s \right] < \infty\\
&\mathbb{E}\left[\int_0^{\tau_n}\left(\frac{\partial f}{\partial p} (s,Z^a_s, \Pi_s) (\rho \sigma_0 + P_s)\right)^2 \ud s\right]\leq   \sup_{t \le T, \ (\zeta,p) \in [-n,n]^2} \left|\frac{\partial f}{\partial p} (t, \zeta, p)\right|^2 2 ( \sup_{t \le T} |P_t|^2 + \rho^2 \sigma^2_0) T < \infty\\
&\mathbb{E}\left[\int_0^{\tau_n} \int_{\R^+} \left|f(s, Z^a_{s-} - u_sz, \Pi_s) - f(s, Z^a_{s-}, \Pi_s) \right| \lambda F_Z(\ud z)  \ud s\right]\leq  \sup_{t \le T, \ (\zeta,p) \in [-n,n]^2} 2\left|f(t,\zeta, p)\right| \lambda T < \infty
\end{align}
which guarantee that $\left\{M_{\tau_n \wedge t}\right\}_{t \in [0,T]}$ is a martingale.

From the equation \eqref{HJB} it holds that for any $a \in [0,1] \times \mathbb{R}$, ${\L}^a f(s,\zeta, p) \geq 0$. Therefore,  for any $n\in \bN$ and $a =(u, \theta) \in \mathcal{A}$, taking the conditional expectation in \eqref{eq:function_f} (written between  $t\wedge \tau_n$ and $T\wedge \tau_n$) we get that
\begin{equation}\label{EXP}
    \mathbb{E}_{t,\zeta, p}\bigl[f(T\wedge \tau_n, Z^a_{T\wedge \tau_n}, \Pi_{T\wedge \tau_n})  \bigr] \geq \mathbb{E}_{t,\zeta, p}\bigl[ f(t\wedge \tau_n, Z^a_{t\wedge \tau_n}, \Pi_{t\wedge \tau_n})  \bigr]
    \end{equation}
where $(t,\zeta, p) \in [0,T]\times \mathbb R^2$.
    Letting $n \rightarrow + \infty$, and using the fact that the process $Z^{a}$ does not have any deterministic jump time it holds that 
    $$f(T\wedge \tau_n, Z^a_{T\wedge \tau_n}, \Pi_{T\wedge \tau_n}) \rightarrow e^{-\eta Z^{a}_T}, \  f(t\wedge \tau_n, Z^a_{t\wedge \tau_n}, \Pi_{t\wedge \tau_n})\rightarrow f(t, Z^a_t, \Pi_t) \ \P -\mbox{a.s.},$$ for 
    any $a =(u, \theta) \in \mathcal{A}$. Condition $(i)$ in the statement of the theorem allows us apply the limit under expectation in  \eqref{EXP}, thus 
    \begin{equation}\label{ver}\mathbb{E}_{t,\zeta, p}\bigl[e^{-\eta Z^{a}_T}\bigr] \geq  f(t,\zeta, p),\end{equation}
    which implies $v(t,\zeta, p) \geq f(t, \zeta, p)$.  By similar computations we can prove that  equality holds in \eqref{ver} when taking the control $\{a^*(t,Z^{a^*}_{t^-},\Pi_t)\}_{t \in [0,T]} \in \mathcal{A}$.
    Consequently, 
   $$v(t,\zeta, p) = \mathbb{E}_{t,\zeta, p}\bigl[e^{-\eta Z^{a^*}_T}\bigr] = f(t, \zeta, p),$$
   which concludes the proof.

\subsection{Proof of Lemma \ref{squareMGM}}

Using the equations \eqref{eq:Lhat} and \eqref{eq:IQ},  we immediately get that  
$$\widehat L^{-1}_T =  e^{- \frac{1}{2} \int_0^T \Pi^2_s \ud s + \int_0^T \Pi_s \ud I^{Q}_s}.$$
Then it holds that 
\begin{align}
  \mathbb{E}^{\Q}[\widehat L^{-2}_T] = &  \mathbb{E}^{\Q}[ e^{-  \int_0^T \Pi^2_s \ud s + \int_0^T 2 \Pi_s \ud I^{Q}_s}] = 
   \mathbb{E}^{\Q}[ e^{-  4\int_0^T  \Pi^2_s \ud s + \int_0^t 2 \Pi_s \ud I^{Q}_s } e^{ 3\int_0^T \Pi^2_s \ud s} ] \\
   & \leq \frac{1}{2} \Big( \mathbb{E}^{\Q} \bigr[ e^{- 8 \int_0^T \Pi_s^2 \ud s + 4 \int_0^T \Pi_s \ud I^{Q}_s}\bigl] + \mathbb{E}^{\Q}[e^{ 6 \int_0^T \Pi^2_s \ud s} ] \Big ) < \infty 
 \end{align}
where in the penultimate inequality we have used that $ab \leq \frac{1}{2}(a^2 + b^2)$ for any $a,b\in \R$ and in the last one we have used that $\{e^{- 8 \int_0^t \Pi_s^2 \ud s + 4 \int_0^t \Pi_s \ud I^{Q}_s} \}_{t \in [0,T]}$ is a positive local martingale, under the measure $\mathbf{Q}$ and the filtration $\bF$, and  thus a supermartingale with finite expectation.

\subsection{Proof of Proposition \ref{uniform}}

We will show that $\sup_{t \in [0,T]}\mathbb{E}[v^{1 + \epsilon}(t, Z^a_t, \Pi_t)] < \infty$, for some $\epsilon >0$. Using the form of the function $v$ (cfr. equation \eqref{v1}) we get that

\begin{align} 
& \mathbb{E}[v^{1 + \epsilon}(t, Z^a_t, \Pi_t)] \leq   A_1 \mathbb{E}\Bigl[e^{-\eta e^{r(T-t)} (1 +\epsilon) Z^{a}_t} e^{-{\frac{1}{2}} (1+\epsilon)\int_t^T \{ \Pi_t e^{-\int_t^s a(l) \ud l} + b_0\mu_0 \int_t^s e^{-\int_v^s a(l) \ud l} \ud v   \} \ud s}\Bigr] \\
& \leq \frac{1}{2} A_1\left(\mathbb{E}\Bigl[e^{-2\eta e^{r(T-t)} (1 +\epsilon) Z^{a}_t}\Bigr] + \mathbb{E}\Bigl[ e^{- (1+\epsilon)\int_t^T \{ \Pi_t e^{-\int_t^s a(l) \ud l} + b_0\mu_0 \int_t^s e^{-\int_v^s a(l) \ud l} \ud v   \} \ud s}\Bigr] \right)\\
& \leq \frac{1}{2} A_1 \left(\mathbb{E}\Bigl[e^{-2\eta e^{r(T-t)} (1 +\epsilon) Z^{a}_t}\Bigr] + A_2 \mathbb{E}\Bigl[ e^{- (1+\epsilon) \Pi_t (T-t) }\Bigr]\right) \\
& \leq \frac{1}{2} A_1 \left(\mathbb{E}\Bigl[e^{-2\eta e^{r(T-t)} (1 +\epsilon) Z^{a}_t}\Bigr] + A_2 \mathbb{E}\Bigl[ e^{ (1+\epsilon) |\Pi_t| T }\Bigr]\right) \label{eq:lastline}
\end{align} 
for some non-negative and constant $A_1, A_2$, where in the first inequality we have used that $ab \leq \frac{1}{2}(a^2 + b^2)$, $a,b\in \R$.  In \eqref{eq:lastline}, the first expectation is finite  because of admissibility of the strategy (see the second condition of \eqref{admissible}). 
The second expectation, instead, is finite because  the process $\Pi$ is Gaussian with mean $m(t) = \Pi_0 e^{-b_0t} + {\mu_0}(1- e^{-b_0t})$, variance $\sigma^2_\Pi(t) = \frac{(P_t + \rho \sigma_0)^2}{2 b_0}(1- e^{-2 b_0t})$.

\subsection{Proof of Proposition \ref{adm_star}}

First, $u^*$ is deterministic process with values in $[0,1]$ and so it is $\bF$-predictable. From \eqref{tetapartial} we get
\begin{equation}\label{stim}|\theta^*_t|^2 \leq c_1(t)\Pi^2_t + c_2(t)
\end{equation}
with 
$$c_1(t)=  \frac{ 2 e^{-2r(T-t)}}{ \sigma^2_1 \eta^2}, \quad c_2(t)= \frac{1}{2} \left ( \frac{ \sup_{s\in [0,T]} P_s + \rho \sigma_0  }{\sigma_1 \eta} e^{-r(T-t)} (T-t)\right)^2$$ 

here we have used  that $P_s$ is a deterministic continuous function, so bounded over $[0,T]$, and that $(a+b)^2 \leq 2(a^2 + b^2)$ for any $a,b\in \R$. 

Next recalling that for any fixed $t \in [0,T]$, $\Pi_t$ has Normal distribution with mean $m(t) = \Pi_0 e^{-b_0t} + {\mu_0}(1- e^{-b_0t})$ and variance $\sigma^2_\Pi(t) = \frac{(P_t + \rho \sigma_0)^2}{2 b_0}(1- e^{-2 b_0t})$, 
we get that
$$\mathbb{E}\Bigl[\int_0^T |\theta^*_t|^2\ud t\Bigr] \leq \int_0^T \big (c_1(t) \mathbb{E}[\Pi^2_t] + c_2(t) \big ) \ud t  < \infty.$$

So the first condition in \eqref{admissible} is satisfied. 

We discuss now the second  inequality in \eqref{admissible}. 
We would like to show that for some  $\epsilon>0$
\begin{align}
\sup_{t \in [0,T]}\mathbb{E}\Bigl[e^{-2\eta e^{r(T-t)} (1 +\epsilon) Z^{a^*}_t}\Bigr]<\infty.
\end{align}
Using the explicit solution of equation \eqref{Za2}, i.e.
$$
dZ^a_t  = R_0 e^{rt} + \int_0^t e^{r(t-s)} \bigl(c-q(u_s) +  \theta_s \sigma_1 \Pi_s \bigl)\,ds +   \int_0^t e^{r(t-s)} \theta_s \sigma_1  \ud I_s -   \int_0^t  \int_{\R^+} e^{r(t-s)} z u_s m(\ud s, \ud z), $$
and independence between the loss process and the financial market, we get that
\begin{align}
&\mathbb{E}\Bigl[e^{-2\eta e^{r(T-t)} (1 +\epsilon) Z^{a^*}_t}\Bigr] \leq \tilde{A} e^{2\eta (1 +\epsilon) e^{rT} q(0) T}\mathbb{E}\Bigl[e^{2\eta (1 +\epsilon) e^{rT}C_T} \Bigr]  \mathbb{E}\Bigl[e^{-2\eta \sigma_1 (1 +\epsilon)\int_0^t e^{r(T-s)} \theta^*_s (\Pi_s\ud s + \ud I_s )} \Bigr],
\end{align}
for some nonnegative constant $\tilde A$. The first expectation is finite as the assumption on the claim size distribution $\mathbb{E}\Bigl[e^{a Z_1} \Bigr] < \infty$, implies in particular that $\mathbb{E}\Bigl[e^{a C_T} \Bigr] < \infty$ for any $a >0$. 

Moreover, using the same trick as in Lemma \ref{squareMGM}, we have that for any $t \in [0,T]$ 
\begin{align}
&\mathbb{E}\Bigl[e^{-2\eta \sigma_1 (1 +\epsilon)\int_0^t e^{r(T-s)} \theta_s^* (\Pi_s\ud s + \ud I_s )} \Bigr] = \mathbb{E}^{\Q}[\widehat L^{-1}_T  e^{-2\eta \sigma_1 (1 +\epsilon)\int_0^t e^{r(T-s)} \theta^*_s \ud I^{Q}_s}] \\
&\le \frac{1}{2}\Bigl(\mathbb{E}^{\Q}[\widehat L^{-2}_T] + \mathbb{E}^{\Q}[e^{ 8 \eta^2 (1+\epsilon)^2 \sigma^2_1\int_0^T e^{2rT}(\theta^*_s)^2 \ud s}] \Bigr),
\end{align}
where, the first equality follows from Girsanov theorem. 
Note that $\mathbb{E}^{\Q}[\widehat L^{-2}_T]$ is finite because of Lemma \ref{squareMGM}. 

Let $k= 8 \eta^2 (1 + \epsilon)^2 \sigma_1^2 e^{2rT}$, by \eqref{stim} we obtain 
$$\mathbb{E}^{\Q}\Bigl[e^{k \int_0^T (\theta_t^*)^2\ud t} \Bigr] \leq \mathbb{E}^{\Q}\Bigl[e^{k \int_0^T (c_1(t)\Pi^2_t + c_2(t)) \ud t }   \Bigr] \leq e^{k \int_0^T  c_2(t) \ud t} \mathbb{E}^{\Q} \Bigl[ e^{ \int_0^T k c_1(t)\Pi^2_t \ud t} \Bigr].$$   
Finally, the expectation on the right hand is finite because 
$k c_1(t) \leq 8 \eta^2 (1 + \epsilon)^2 \sigma_1^2 e^{2rT} \times  \frac{ 2 }{ \sigma^2_1 \eta^2} = K$, for any $t \in [0,T]$. Summarizing, we obtain that
\begin{align}
& \sup_{t \in [0,T]}\mathbb{E}\Bigl[e^{-2\eta e^{r(T-t)} (1 +\epsilon) Z^{a^*}_t}\Bigr] \leq \\
& \frac{1}{2} \tilde{A} e^{2\eta (1 +\epsilon) e^{rT} q(0) T}\mathbb{E}\Bigl[e^{2\eta (1 +\epsilon) e^{rT}C_T} \Bigr] 
\Bigl(\mathbb{E}^{\Q}[\widehat L^{-2}_T] + e^{k \int_0^T  c_2(t) \ud t} \mathbb{E}^{\Q} \Bigl[ e^{ K \int_0^T \Pi^2_t \ud t} \Bigr] \Bigr)< \infty,
\end{align}
and this concludes the proof.

\subsection{Proof of proposition \ref{prop:adm2}}
Recall the representation of the filter given by 
$$\Pi_t = \Pi_0  e^{-\int_0^t a(l) \ud l} + b_0\mu_0 \int_0^t e^{-\int_v^t   a(l) \ud l} \ud v +
\int_0^t \left(P_v +\rho \sigma_0 \right) e^{- \int_v^t a(l) \ud l} \ud I^Q_v, $$
(see also Remark \ref{rapp}); hence under $\Q$, $\Pi_t$ has Gaussian distribution with variance $\sigma^{Q,2}_\Pi(t)$ that satisfies for any $t\in [0,T]$ 
$$\sigma^{Q,2}_\Pi(t) = \int_0^t \left(P_v +\rho \sigma_0 \right)^2 e^{- 2\int_v^t a(l) \ud l} \ud v \leq \frac{(\bar P + \rho\sigma_0)^2}{2 (b_0 + \rho\sigma_0)} (1- e^{-2 (b_0 + \rho\sigma_0)t}) \leq \frac{(\bar P + \rho\sigma_0)^2}{2 (b_0 + \rho\sigma_0)}.$$ 
Therefore, arguing as in the proof of Lemma \ref{Nov} we get the claimed result.
 

 \end{document}